\documentclass[twocolumn,prb,preprintnumbers,amsmath,amssymb,floatfix,showpacs]{revtex4}

\usepackage{graphicx}
\usepackage{subfigure}
\usepackage{dcolumn}

\begin{document}

\title{First Principles Study of the Electronic and Vibrational Properties of LiNbO$_2$}
\author{Erik R. Ylvisaker}
\author{Warren E. Pickett}
\affiliation{Department of Physics, University of California, Davis, California, 95616}


%
\begin{abstract}

In the layered transition metal oxide LiNbO$_2$ the Nb$^{3+}$ ($4d^2$) ion is trigonal-prismatically coordinated with O ions, with the resulting crystal field leading to a single band system for low energy properties.  A tight-binding representation shows that intraplanar second neighbor hopping $t_2 = 100$ meV dominates the first neighbor interaction ($t_1 = 64$ meV).  The first and third neighbor couplings are strongly modified by oxygen displacements of the symmetric Raman-active vibrational mode, and electron-phonon coupling to this motion may provide the coupling mechanism for superconductivity in Li-deficient samples (where $T_c \approx 5$ K).  We calculate all zone-center phonon modes, identify infrared (IR) and Raman active modes, and report LO-TO splitting of the IR modes.  The Born effective charges for the metal ions are found to have considerable anisotropy reflecting the degree to which the ions participate in interlayer coupling and covalent bonding.  Insight into the microscopic origin of the valence band density, composed of Nb $d_{z^2}$ states with some mixing of O $2p$ states, is obtained from examining Wannier functions for these bands.

\end{abstract}
%

\maketitle


%
\section{Introduction}

The quasi-two-dimensional (2D) compound LiNbO$_2$ has become of 
interest in condensed matter studies in no small part
due to the existence of superconductivity\cite{Geselbracht-Nature} 
in the de-lithiated phase Li$_x$NbO$_2$.
This system consists of a triangular lattice of transition
metal (niobium) ions separated by layers of O ions from `blocking'
(Li) layers, thus possessing structural
similarities to high temperature 
superconducting cuprates.  
Important differences from the cuprates include the presence of a $4d$ rather than $3d$ ion, 
and of course the triangular rather than square lattice.  
The trigonal prismatic coordination of Nb with six O atoms causes the valence band to be composed solely
of Nb d$_{z^2}$ states.  Other d states are well separated by a 2 eV gap, leaving only a single band per formula unit required for analysis.  The comparative structural simplicity of 
LiNbO$_2$ makes it attractive for theoretical 
studies, especially as it requires modest computational time to 
characterize its properties.  

This alkaliniobate is also of interest due to the increasing attention garnered by niobates in recent years.\cite{simon}  The three-dimensional LiNbO$_3$ is an important and well studied ferroelectric material.\cite{linbo3}  Superconductivity at T$_c \sim$ 20 K has been reported by a few groups\cite{banbo1,banbo2,banbo3}
in the barium niobate system.  The phase has not been positively identified but it has been suggested to be a nonstoichiometric BaNbO$_x$ cubic perovskite.  Superconductivity in the 5-6 K range has been reported in the
Li-intercalated layered perovskite system Li$_x$AB$_2$Na$_{n-3}$Nb$_n$O$_{3n+1}$ [A = K, Rb, Cs;
B = Ca, Sr, Ba; $n$ = 3 and 4].\cite{takano,kato}  This system has clear structural motifs
in common with the peculiar Nb$_{12}$O$_{29}$ system, where the `single O$^{2-}$ deficiency'
(with respect to insulating Nb$_2$O$_5$ = Nb$_{12}$O$_{30}$)
releases two carriers into the cell; one localizes and is magnetic while the other is itinerant
and results in conducting behavior.\cite{waldron,andersen04} 

While there is good agreement as to structure type and lattice parameters of the stoichiometric phase LiNbO$_2$, there is some disagreement in the literature as to its electronic and magnetic properties.  Originally it was reported as being paramagnetic with strong field dependence,\cite{Meyer} however calculations predict it to be a diamagnetic semiconductor. \cite{Burdett}  Other experimental work suggests that it is diamagnetic, and strongly field-dependent paramagnetic properties are the result of impurities.\cite{Geselbract}  The values of the band gap are also in mild dispute.  Extended H\"uckel calculations with a single isolated NbO$_2$ layer give a bandgap value of 1.4  eV, \cite{Burdett} and full potential linear muffin tin orbital calculations with the full crystal structure of LiNbO$_2$ give 1.5 eV.\cite{Novikov94-2}  Measured optical reflectance shows an onset at $\sim 2$ eV, which is suggested to be a direct bandgap responsible for the burgundy-red color of LiNbO$_2$.\cite{Geselbract}  Band structure calculations given here and by Novikov {\em et al.}\cite{Novikov94-2} are in agreement that LiNbO$_2$ is a direct gap semiconductor.

Geselbracht, Richardson and Stacy\cite{Geselbracht-Nature} first reported that the de-lithiated phases Li$_x$NbO$_2$ 
with $x = 0.45$ and $x=0.50$ support superconductivity at temperatures below
5.5 K and 5 K, respectively.  Moshopoulou, Bordet and Capponi\cite{Mosh} later
reported samples with $x$ ranging from 0.69 to 0.79 showing the onset of the Meissner effect at
5.5 K, but samples with $x=0.84$ and above do not exhibit any superconducting transition down
to 2 K.
Superconductivity has also been reported in hydrogen-inserted H$_x$LiNbO$_2$, in which both
samples reported had the same $T_c = 5.5$ K for $x = 0.3$ and $x = 0.5$. \cite{Kumada}

Removal of Li has the effect
of adding holes to the conduction band made up of Nb $d_{z^2}$ states.  
As mentioned above, there does not appear to be any notable 
dependence\cite{Geselbracht-Nature,Mosh} of $T_c$ on the 
concentration of Li in the range $0.45 < x < 0.8$.  Such concentration independence has also been
observed in the 2D electron-doped system\cite{zrncl,zrncl2} Li$_x$ZrNCl. For a phonon-paired superconductor
with cylindrical Fermi surfaces this behavior is understood in terms of phase space restrictions
on phonon scattering processes\cite{role2D} in 2D.
The independence of $T_c$ on carrier concentration in this system is quite
surprising however, since the density of states (DOS) plots presented by Novikov {\em et al}.\cite{Novikov94-2} and here varies strongly with energy, whereas parabolic bands lead to a constant
DOS in 2D.  There has also been some 
work done investigating the ion mobility of LiNbO$_2$ to 
assess its potential usefulness as a battery material.\cite{Geselbract,McDowell}

In this work we study the electronic structure and bonding of stoichiometric LiNbO$_2$ using density functional theory in the local density approximation (LDA).  Density functional perturbation theory is used to obtain zone center phonons and Born effective charges. The structure contains a single parameter $z$ that specifies the position of the O layers relative to the Li and Nb layers, and which seems to be of particular importance in understanding the electronic bonding and electron-phonon coupling in this compound.  We investigate the corresponding Raman active vibrational mode (beating oscillation of the O planes) and the effect this has on the band structure.  We obtain a tight-binding (TB) model that involves many neighbors and helps one to understand how this variation affects interactions between the Nb atoms.  Many aspects of the  electronic structure, such as chemical bonding and effective charges, can be seen to be interrelated by examining the Wannier functions of the valence bands.

%
\section{Structure}

Structural information of LiNbO$_2$ has been reported by several experimental
groups, leading to the assignment of space group $P6_3/mmc$ (No. 194), with Li occupying 
sites 2a (0,0,0) with symmetry ${\overline 3}m$, Nb occupying 
sites 2d $(\frac23, \frac13, \frac14)$ with symmetry ${\overline 6}m2$, and O occupying
sites 4f $(\frac13, \frac23, z)$ with symmetry $3m$.  The O sites include 
an internal parameter $z$ specifying their height relative to the Li layer along the $c$-axis.  
The Li sites are centers of inversion; the Nb sites have z-reflection symmetry
in addition to the $3m$ triangular symmetry within the layer but do not lie
at centers of inversion.  

There is good experimental agreement on lattice parameters, $a = 2.90 \hbox{\AA}$, $c = 10.46 \hbox{\AA}$.\cite{Mosh, Geselbract, Meyer}  The distance $a$ between Nb atoms is quite close to the nearest neighbor distance of 2.86~\AA\, in elemental bcc Nb, suggesting that direct Nb-Nb coupling should be kept in mind.  However, Nb$^{3+}$ $4d$ orbitals will be smaller than in neutral Nb.  In later sections, we will discuss details of the Nb-O interactions, and show that the electronic properties of LiNbO$_2$ are very sensitive to changes in the Nb-O distance.  The vertical distance between Nb and O planes can be calculated by $(\frac14- z)c$, and the distance between Nb and O atoms is given by $\sqrt{\frac29a^2 + (\frac14-z)^2c^2}$.

Some disagreement exists about the value of $z$.  A larger value of $z$
indicates the O layers are closer to the Nb layers.  Meyer and Hoppe\cite{Meyer}
found $z = 0.1263$, Moshopoulou, Bordet and Capponi\cite{Mosh} report $z = 0.128$,
Geselbracht, Stacy, Garcia, Slibernagel and Kwei\cite{Geselbract} report $z = 0.12478$, 
Tyutyunnik, Zubkov, Pereliaev and Kar'kin\cite{Tyut} report 
$z = 0.1293$.  These variations amount
to a difference in O layer position of about 0.05\AA.

Optimization of unit cell geometry within LDA leads to $a = 2.847  \hbox{\AA}$,
$c = 10.193$ \AA, $z = 0.1211$.  This differs from the measured value by $\Delta a \approx -1.8\%$, 
$\Delta c \approx -2.5\%$, for a total volume discrepancy of $\Delta V/V \approx -6\%$.  
The calculated overbinding of the unit cell is somewhat larger than is typical in LDA calculations,
perhaps due to the low (practically zero) valence electron density in the Li layer. 
In this regard, we note that our Nb pseudopotential does not include the $4p$ states in the
valence bands, which may have some effect on structural properties.
Optimization with respect to $z$, holding the lattice parameters $a$ and $c$ fixed at experimental 
values gives $z = 0.125$, which is very close to the value of Geselbracht {\em et al}.\cite{Geselbract}
\begin{figure}[t]
\includegraphics[width=0.45\textwidth]{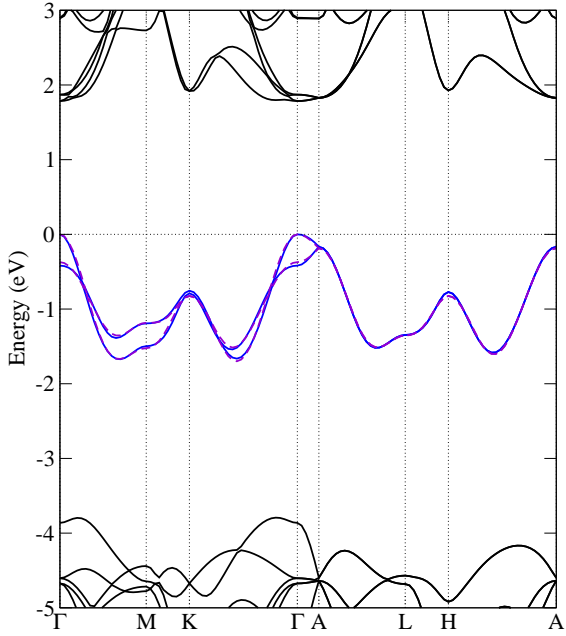}
\caption{(Color online) Band structure for LiNbO$_2$ with O height $z = 0.1263$ typical of the slightly
scattered experimental results.  Two Nb bands (one per Nb layer) lie within a 5.5 eV gap
between the O $2p$ bands below and the remaining Nb $4d$ bands above. 
The tight binding fit of Table \ref{tbl:TightBinding} is shown with a dashed line.}
\label{fig:Band}
\end{figure}
\begin{figure}[thb]
  \includegraphics[width=0.45\textwidth]{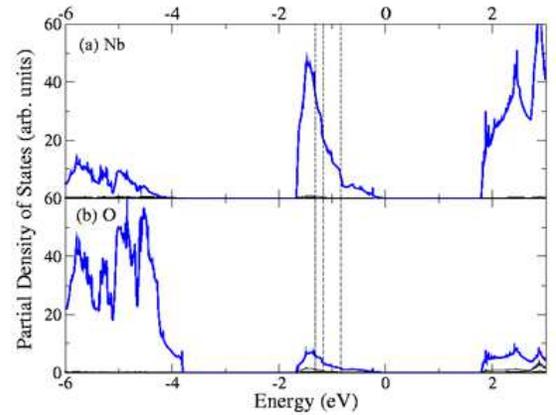}
  \caption{(Color online) Density of states for Nb and O atoms.  For LiNbO$_2$, states are 
  	occupied up to energy $E = 0$.  The dominant states in the valence band for Nb
  	are $4d$-orbitals, and for O they are $2p$-orbitals.  Dashed vertical lines are included to indicate
  	the Fermi energy in the rigid band picture for Li$_x$NbO$_2$ for 
  	concentrations $x = 0$ (half-filled band), $x = 0.4$, $x = 0.8$ from left to right.
  The last two values of $x$ roughly bracket the concentrations where superconductivity
  is observed. 
  	}
  \label{fig:ProjectedDOS}
\end{figure}
%

%
\section{Electronic Structure Calculations}

The present results have been obtained through the use of the {\sc abinit} code.\cite{abinit}  The lattice constants and the parameter $z$ given by Meyer and Hoppe\cite{Meyer} were used in the calculations in this section.  Pseudopotentials used were LDA, Troullier-Martins type,\cite{Troullier} generated using the Perdew-Wang LDA exchange-correlation functional.\cite{Perdew-Wang}  A self-consistent density was generated with 36 irreducible $\mathbf{k}$-points, which was then used to calculate energies at 116 $\mathbf{k}$-points along high symmetry directions to plot the band structure.  The kinetic energy cutoff for the planewave basis was set to 40 Hartrees.

Figure \ref{fig:Band} of the important region of the band structure shows the top of the O $2p$ bands, a 2 eV gap
to two Nb $4d$ bands which are the highest occupied bands, and another 2 eV gap to the conduction bands.  The band gap, direct at the $\Gamma$ point, is 1.9 eV, somewhat higher than the 1.5 eV reported by Novikov {\em et al}. (also at the $\Gamma$ point) using the full potential linear muffin-tin orbital method (FLMTO) with the same structural parameters.\cite{Novikov94-2} LDA calculations often underestimate band gaps, so we conclude that it is most likely the experimental band gap is somewhat larger than 2 eV. Measured optical absorption suggests that the band gap should be around 2 eV,\cite{Geselbract} which is consistent with the reported dark reddish color.\cite{Geselbract, Meyer}  

Projected density of states (DOS) calculations, presented in Fig. \ref{fig:ProjectedDOS} were done using the tetrahedron method projecting to atomic spheres with radii of R(Li) = 1.7 a.u., R(Nb) = R(O) = 2.0 a.u.  These calculations confirm that the uppermost valence bands are largely Nb $4d$ character, with roughly 15\% contribution from O $2p$ states, and virtually none from Li states. These projected DOSs are consistent with those calculated by 
Novikov {\em et al}.\cite{Novikov94-2}  The trigonal crystal field splitting of the Nb $3d$ states results in the $d_{z^2}$ orbital energy centered 4 eV below the other Nb $4d$ states (two $e_g$ doublets, based on \{$xz,yz$\} and \{$x^2-y^2,xy$\}).  The unit cell contains two symmetry-related Nb layers which result in the presence of two Nb $d_{z^2}$ bands in the band plot.  The splitting of the bands along $\Gamma$-M-K-$\Gamma$ reflects the strength of interaction between layers, as discussed below.

As a result of the large trigonal crystal field splitting, the Nb $d_{z^2}$ 
state forms a single band, triangular lattice system of the type that is attracting renewed
interest.  The calculated DOS in Fig. \ref{fig:ProjectedDOS} has only a
vague resemblance to the nearest neighbor TB model
(see for example Honercamp\cite{honercamp}) making it evident that longer range hopping must
be important.  We address this question in detail in section \ref{Sec:TB}.   

This general electronic structure, when hole-doped, is rather remarkable.  It is a {\it bona fide}
representation of a single band system on a triangular lattice, moreover the band is well
isolated from other bands.  Although single band models are widely used to investigate
concepts and to model many-body effects, there are very few actual realizations in real
solids.  The value of the effective Coulomb on-site repulsion $U$ for Nb$^{3+}$
is not established, but with $W < 1.75$ eV it will not require a very large value
to lead to correlated behavior.  The hole-doped system Li$_x$NbO$_2$ provides a unique platform for the detailed comparison of single band models with experimental data, which for thermodynamic properties and for spectroscopic
data below 2 eV should involve only the single active band.

\begin{figure*}[tb]
\begin{centering}
\includegraphics[width=0.9\textwidth]{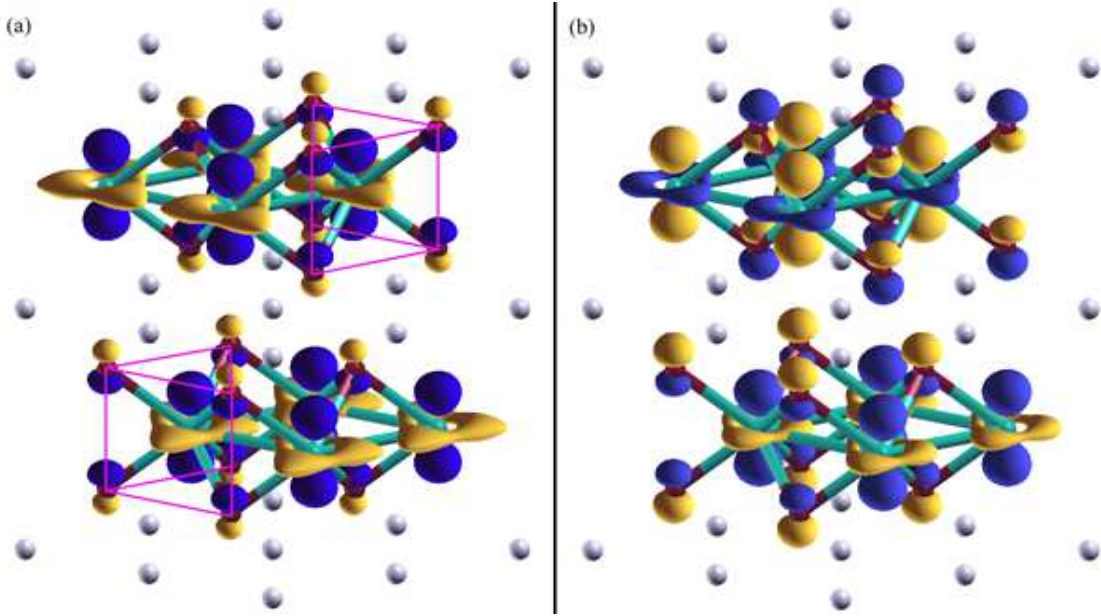}
\caption{(Color online) Isosurface plot of the $\Gamma$-pt wavefunctions for LiNbO$_2$ for
(a) the lower of the pair of valence bands, and (b) the upper valence band, made with 
XCrysDen.\cite{XCrysDen}  Both wavefunctions are generated with
the same isosurface value.  
Nb atoms are turquoise (grey), O atoms are dark red (dark gray), Li are light gray.  
Yellow and blue (light and dark) indicate opposite signs of the wavefunction.
The $z^2$ symmetry around the Nb atom is apparent as well as
the contribution to the bonding made by the O atoms.  Trigonal prisms with O atoms at the corners
and Nb atoms in the center have been outlined as a guide to the eye.  
\label{fig:wavef-17-18} }
\end{centering}
\end{figure*}

%
\section{Zone Center Wave functions}

We can examine Kohn-Sham wavefunctions to gain insight into the electronic structure.  Shown in Fig. \ref{fig:wavef-17-18} are isosurface plots of the $\Gamma$ wavefunctions for the lower and upper valence bands.  Both plots reveal the $d_{z^2}$-character of the bands on the Nb atoms, as well as $2p_z$ bonding contributions from the O atoms.  For the upper band, the $p$-lobes have more weight away from the sandwiched Nb layer, indicating the weight of bonding with O is somewhat decreased.  The change in sign between the two Nb layers in Fig. \ref{fig:wavef-17-18}b indicates the presence of a nodal plane in the Li layer arising from the wavefunction's antibonding character (with respect to interlayer coupling).

The presence of an interesting triangular shape of the wavefunction around the waist of the Nb atoms is more unexpected, being noticeably larger for the bonding function.  The corners of these triangles do not point toward nearest neighbors, instead they point toward interstitial holes in the Nb layer which are not directly sandwiched between O atoms.  This shape reflects either three-center Nb-Nb bonding, which would result in an increased density in this region, or antibonding character of direct Nb-Nb interactions that would decrease the density along the directions connecting Nb atoms.  The character and origin of this triangular `waist' around Nb atoms are clarified in the following section.

%
\section{Wannier Function}
\label{Sec:Wannier}

\begin{figure}[th]
\includegraphics[width=0.45\textwidth]{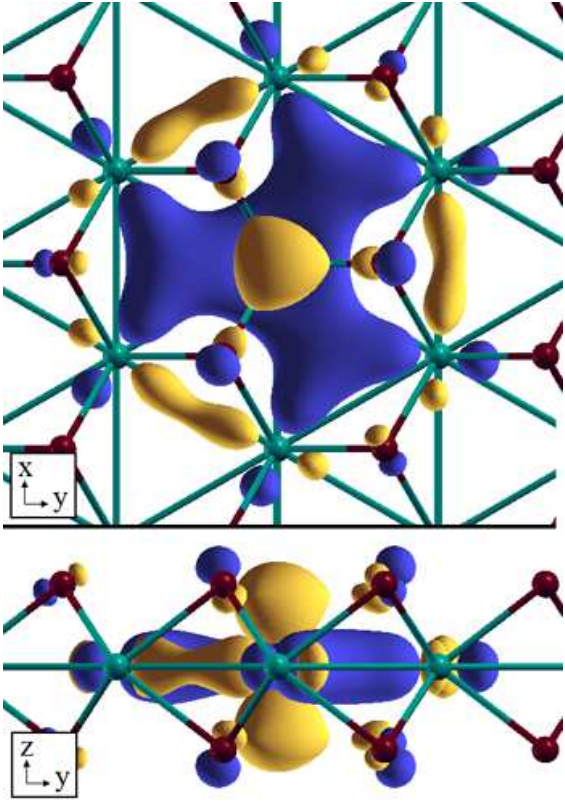}
\caption{(Color online) Isosurface plot of the Wannier function for LiNbO$_2$, made with
XCrysDen.\cite{XCrysDen}  
Niobium atoms are turquoise (gray), O atoms are dark red (dark gray), Li not pictured.  
Yellow and blue indicate opposite signs of the WF.
The $z^2$ symmetry around the central atom is apparent, as well as the 
$xy$/$x^2-y^2$ character around nearest neighbors, which maintains the orthogonality
of WFs centered on different atoms. \label{fig:WF} }
\end{figure}

A single isolated band also provides an unusually clean system for performing the 
transformation from reciprocal space (bands) to real space (bonds) to look at 
character and strength of interatomic interactions from a local viewpoint.
The straightforward way of doing this is to generate Wannier functions (WFs) defined by
\begin{equation}
	|\mathbf{R}m\rangle = N_\mathbf{k}^{-1/2} 
		\sum_{\mathbf{k},n} U_{mn}^{\mathbf{k}} 
		e^{-i\mathbf{k}\cdot\mathbf{R}} |\mathbf{k}n\rangle
\end{equation}
where $U^{\mathbf{k}}$ can be any unitary matrix, and $N_\mathbf{k}$ is the number of k-points used in the summation.  The orthonormal Bloch states are denoted by $|\mathbf{k}n\rangle$.  If there were only one NbO$_2$ layer per cell, hence only one band, $U(\mathbf{k})$ would be simply a complex number of unit modulus.  Here it is a $2\times 2$ matrix in the band space, which gives rise to very modest extra complexity. These WFs display orthonormality
$\langle\mathbf{R}m | \mathbf{R}'m'\rangle = \delta_{\mathbf{R},\mathbf{R}'}\delta_{m,m'}$. Once a choice for $U^\mathbf{k}$ is made, this approach generates a basis set of identical WFs for each point in the Bravais lattice of the system being studied. 

We have chosen the unitary matrix $U$ such that $U = M(M^\dagger M)^{-1/2}$ with $M_{mn} = \langle \mathbf{k}n | S_m \rangle$ where $|S_m\rangle$ are a set of atom-centered functions with the desired $d_{z^2}$symmetry.  This technique has been used previously in the study of magnetic ordering in cuprates.\cite{Wei-PRL}  Choosing $|S_m\rangle$ as hydrogenic 4$d_{z^2}$ orbitals centered on the Nb atoms for the two entangled valence bands gives a pair of identical WFs for each Nb bilayer in the unit cell, based on the Nb $d_z^2$ symmetry that was also observed in the $\Gamma$-point wavefunction. An $8 \times 8 \times 4$ mesh totaling 256 k-points was used to generate the WF shown in Fig. \ref{fig:WF}.  Increasing the size of the mesh does not change the visual appearance of the WF.

\begin{figure}[t]
\includegraphics[width=0.45\textwidth]{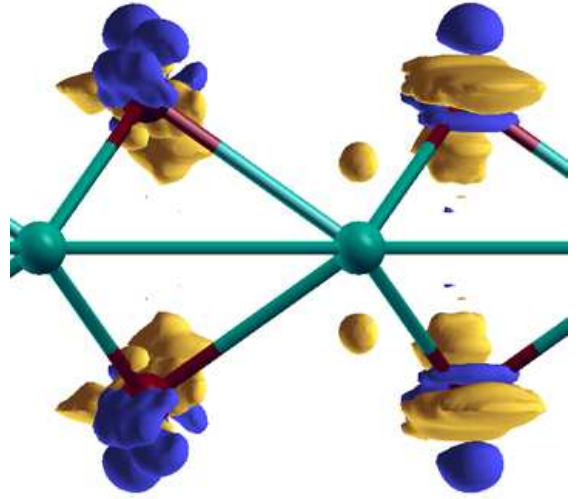}
\caption{(Color online) Isosurface plot of the difference $|W_{z=0.1359}(\mathbf{r})| - |W_{z=0.1263}(\mathbf{r})|$, looking along the [010] cartesian direction.  Yellow (light gray) is the positive isosurface and blue (dark gray) is negative, with Nb atoms in turquoise (gray) and O atoms in dark red (dark gray).  As the O atoms are moved closer to the Nb layer, there is an increase in density above and below the Nb atom; in addition the contributions from the O-p orbitals appear to tilt more towards the central Nb atom.
\label{fig:WF1359diff}}
\end{figure}

The WF naturally displays the prominent lobe along the $\pm \hat z$ axis
that is characteristic of the $3z^2 - r^2$ function.  It also contains
$p_{\sigma}$ contributions from the nearest O ions, as well as smaller
but still clear $2p$ amplitude on the second neighbor O ions that is 
approximately $p_{\pi}$.  Here $p_{\sigma}$
denotes the combination of $p_x, p_y, p_z$ functions oriented toward
the reference Nb atom, while $p_{\pi}$ is perpendicular.  The 
$p_{\sigma}$ lobe nearest the Nb ion has
the same sign as the $3z^2 - r^2$ lobes along the $\hat z$ axis.

The in-plane part of the WF reveals more about the bonding in LiNbO$_2$.
The ring-structure of the $3z^2 - r^2$ function in the $x-y$ plane
connects with combinations of $xy$ and $x^2 - y^2$ functions on the
three neighboring Nb ions to form a substantial in-plane windmill
structure with three paddles, which are oriented between the $z$-projections
of the neighboring O ions.  A ``$3z^2 - r^2$'' symmetry WF has the
full D$_{3h}$ symmetry of the Nb site, which allows a contribution of
an appropriate combination of $x^2 - y^2$ and $xy$ atomic orbitals, and
the windmill structure reflects the bonding of such functions on 
neighboring Nb ions.  Thus the $4d$ character is not entirely $d_{z^2}$,
which clarifies the threefold symmetric shape around the Nb waist in
the $\mathbf{k}=0$ wavefunction discussed in the previous section.

In addition, there is a `hotdog' structure of opposite sign connecting
pairs of nearest-neighbor Nb ions.  Of the two types of Nb-Nb pairs, these hotdogs lie nearest to the first neighbor O ions.
Small nearest-Nb ``backbonds'' in the $x-y$
plane of both the windmill arms and the hotdogs are also evident.
As we note below, the WF contains information about the Nb-Nb hopping
parameters.  The more distant parts of the WF function are a
direct reflection of the existence of such longer range interaction.

This Nb 4d$_{z^2}$ WF is similar to that for Ta 5d$_{z^2}$ in isostructural 
TaSe$_2$.\cite{Wei-arxiv}   A difference is that the nearest neighbor O $2p$ orbitals contribute
more significantly to the bonding in LiNbO$_2$, giving a larger nearest neighbor hopping.  
The d$_{z^2}$ lobes above and below the transition metal appear to be somewhat more pronounced 
in TaSe$_2$.  We also observe
changes toward such appearance in this Nb WF as the O layer height ($z$) is increased, forcing the O atoms closer
to the Nb layers and increasing $t_1$.  Given in Fig. \ref{fig:WF1359diff} is a plot
of the {\em difference} in the magnitudes of the WFs taken for the two different $z$ 
values of 0.1359 and 0.1263 (O displacements of roughly 0.1~\AA.)  
There is an increase in the density of the d$_{z^2}$ lobes above and below the transition metal
as well as a tilting of the O orbitals toward the central site, as the Nb-O distance is reduced.  
Correlating this with the tight binding results from the next section 
suggests there is an increase in the role the O atoms play in the hopping to nearest neighbors as
the Nb-O distance is decreased.  
The opposite effect occurs when $z = 0.1167$; the O orbitals tilt to become oriented more 
along the z-direction, indicating the O atoms are interacting less strongly with Nb.

%
\section{Tight Binding}
\label{Sec:TB}

\begin{figure}[tbh]
\includegraphics[width=0.45\textwidth]{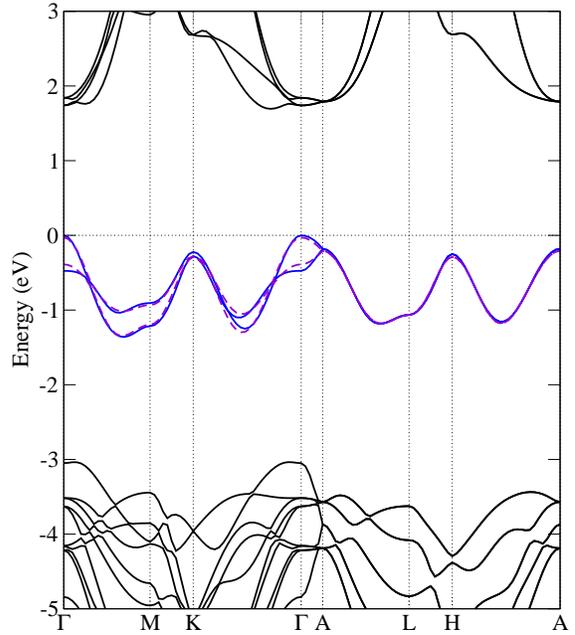}
\caption{(Color online) Band structure for O layer position $z = 0.1167$, a larger Nb-O layer separation
by 0.1~\AA~compared to the bands in Fig. \ref{fig:Band}.  The band width is 25\% smaller,
and band gap is significantly less, compared to the experimental structure ($z = 0.1263$).}
\label{fig:Band1167}
\end{figure}
\begin{figure}[tbh]
\includegraphics[width=0.45\textwidth]{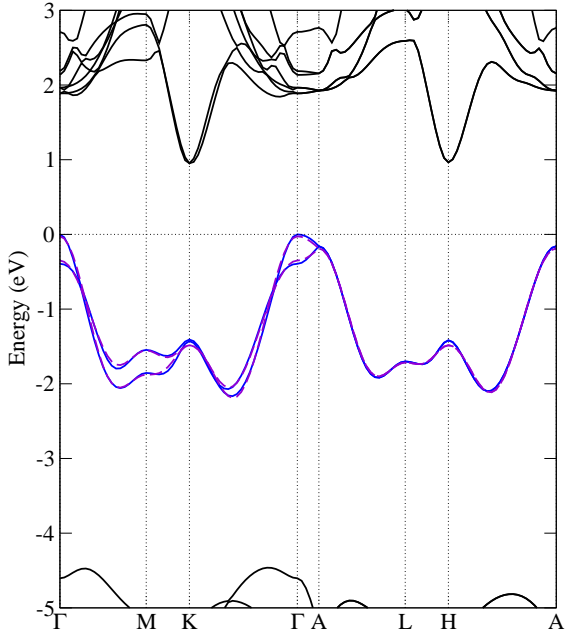}
\caption{(Color online) Band structure for O layer position $z = 0.1359$, about 0.1~\AA~closer to the
Nb layer than in experiment (Fig. \ref{fig:Band}).  The bandwidth is 15\% larger than for the experimental 
structure, and the peaks at M and K in the valence band are similar to the band 
structure calculation of Li$_{0.5}$NbO$_2$ shown by Novikov {\em et al}.,\cite{Novikov94-2} suggesting that the underlying electronic structure of Li$_x$NbO$_2$ is determined by the position of O layers 
rather than the Li concentration.}
\label{fig:Band1359}
\end{figure}

\begin{table*}[th]
\begin{centering}
\begin{tabular}{cccccccccccccc}

\hline \mbox{$z$}& \mbox{$d_{Nb-O}$} 	
							     & \mbox{Bandwidth}	& Band Gap
							     					 	   & \mbox{$t_1$}  
							     					    			& \mbox{$t_2$}  
							     								    		& \mbox{$t_3$}  
							     					  					  			& \mbox{$t_{4}$}   
							     					    									& \mbox{$t_5$}  
							     					   	 										& \mbox{$t^\perp_1$}
							     					   	 											& \mbox{$t^\perp_2$}	
							     					   	 												& \mbox{$t^\perp_3$}	
							     					   	 												& \mbox{$t^\perp_4$}	
							     					   	 												& \mbox{$t^\perp_5$}	\\
\hline \hline 

0.1167			&	2.178			& 1.36		&  1.69	&  19	& $\mathbf{102}$ 	& 10  & -10  & 6  & 18 & -23 & -5 & -7 & -1	\\
0.1263	  		&	2.116			& 1.67		&  1.78	&  64	& $\mathbf{100}$    & 33  & -13  & 8  & 18 & -21 & -5 & -6 & -8	\\
0.1359			&	2.056			& 2.17		&  1.89(0.95) & $\mathbf{122}$  	 & 94   & 56  & -17  & 9 & 18 & -19 & -3 & -6 & -8 \\
\hline
2H-TaSe$_2$		&	
		& 1.9		&   	&  38	& $\mathbf{115}$    &    &    &   & 29 & 23    	\\

\hline

\end{tabular}
\caption[Tight binding parameters]
{Tight binding hopping parameters for three values of the O internal structural parameter $z$.
Hopping parameters are reported in meV; bandwidths and band gaps are reported in eV.  Distance between Nb and
O atoms are reported in angstroms.  The largest hopping coefficients
are in boldface.  Squeezing the O layers closer to the Nb layers ($z=0.1359$) results in 
a dramatic increase in $t_1$, making it larger than $t_2$, and a large change in $t_3$
as well.  For $z=0.1359$, the band gap is indirect.  The direct gap at $\Gamma$ is listed followed by the indirect gap in parentheses.  See figure \ref{fig:Band1359}.  $t^\perp_3$ is slightly refined, to make the splitting at $\Gamma$ agree with the DFT
result better.}
\label{tbl:TightBinding}
\end{centering}
\end{table*}

The valence bands seen in Fig. \ref{fig:Band} are well isolated from bands above and below.  These bands contain the superconducting electrons in de-lithiated Li$_x$NbO$_2$, so a simplified tight binding fit using only these
bands would be useful in doing studies on simple models.
The TB model to fit two bands corresponding to identical Nb d$_{z^2}$-O p hybridized
Wannier functions on the two Nb atoms per cell gives dispersion relations from the roots of 
\begin{equation}
0 = 
\left|
  \begin{array}{ll}
    \varepsilon_\mathbf{k}^\parallel -  \varepsilon_\mathbf{k} 
    				& \varepsilon_\mathbf{k}^\perp      \\
    \varepsilon_\mathbf{k}^\perp*   
    				& \varepsilon_\mathbf{k}^\parallel - \varepsilon_\mathbf{k}  
  \end{array}
\right| \Longrightarrow \varepsilon_\mathbf{k} =  \varepsilon_\mathbf{k}^\parallel 
				\pm | \varepsilon_\mathbf{k}^\perp |
\end{equation}
where $\varepsilon_\mathbf{k}^\parallel$ arises from the hopping processes within
a plane of atoms and $\varepsilon_\mathbf{k}^\perp$ from hopping between planes.
Within the plane of the triangular lattice, the nearest neighbor
dispersion for the six nearest neighbors is
%
%
\begin{equation}
\varepsilon_\mathbf{k}^\parallel = 2t \left[ 
	\cos(k_xa) + 2 \cos \left( \frac{1}{2}k_x a \right) 
			 	  \cos \left( \frac{\sqrt{3}}{2} k_y a \right)
	\right].
\end{equation}
We have found that a good fit to the Nb $d_{z^2}$ bands requires several neighbors.
Third and fifth neighbors in the plane can be included with this form by
noticing odd neighbors just require a larger lattice constant.  Second neighbors require 
a larger lattice constant and a 30$^\circ$ rotation.  There are twelve fourth neighbors, which can be
treated as two lattices with separate rotations.  

First and second neighbors in adjacent layers can be included with terms of the form
%
%
\begin{eqnarray}
  	\varepsilon_\mathbf{k}^\perp = 
		2 \cos \frac{k_z c}2 
		\left\{ 
			2 t^\perp_1 \left[
				\cos \left(\frac{k_xa}2 \right) e^{\frac{ik_y a}{2\sqrt{3}}} + e^{\frac{i k_y a}{\sqrt{3}}} \right] + \right. \nonumber \\
	 \left. 2 t^\perp_2 \left[
	 			\cos \left(k_xa \right)         e^{-\frac{ik_ya}{ \sqrt{3}}} + e^{\frac{-i2k_y a}{\sqrt{3}}} \right]
		\right\}.
	\label{Eperp}
\end{eqnarray}
It is worth noting that on the zone surface (where $k_z = \pm\frac{\pi}{c}$), all
terms in $\varepsilon_\mathbf{k}^\perp$, including further neighbors neglected from 
Eq. (\ref{Eperp}) are identically zero,
resulting in exact degeneracy of the two valence bands at these zone
surfaces.

Using the Wannier basis given in the previous section, the hopping parameters can be
{\it directly calculated} (rather than fit) as matrix elements of the LDA Hamiltonian:
\begin{subequations}
	\label{eq:TBHop}
	\begin{equation}
		\label{eq:TBSingleWFHop}
		t_{|R|,m} = \langle\mathbf{R}m|\hat{H}|\mathbf{0}m\rangle, \quad m = 1,2;					
	\end{equation}
	\begin{equation}
		\label{eq:TBInterWFHop}
		t_{|R|,m,m'}^\perp = \langle\mathbf{R}m'|\hat{H}|\mathbf{0}m\rangle.
	\end{equation}
\end{subequations}
Using the definition of the WF along with orthogonality of the eigenfunctions,
Eqs. (\ref{eq:TBHop}) reduce to 
\begin{subequations}
	\label{eq:TBHopSimplified}
	\begin{equation}
		\label{eq:TBSingleWFHopSimplified}
		t_{|R|,m} = N_{\mathbf{k}}^{-1} \sum_{\mathbf{k}n} \varepsilon_{\mathbf{k}n} |U^{\mathbf{k}}_{mn}|^2 e^{i\mathbf{k}\cdot\mathbf{R}}
	\end{equation}
	\begin{equation}
		\label{eq:TBInterWFHopSimplified}
		t_{|R|,m,m'}^\perp = N_{\mathbf{k}}^{-1} \sum_{\mathbf{k}n} \varepsilon_{\mathbf{k}n} 
							{(U^{\mathbf{k}}_{m'n})^* } U^{\mathbf{k}}_{mn} e^{i\mathbf{k}\cdot\mathbf{R}}.
	\end{equation}
\end{subequations}
where $\mathbf{R}$ is the difference in centers of the WFs.
Deviation of a perfect representation of the bands arises only from approximation of the
summation in Eq. (6) or neglect of further neighbors.

The results of the tight-binding calculation are presented in Table \ref{tbl:TightBinding},
as calculated from a $12 \times 12 \times 6$ Monkhorst-Pack grid of $\mathbf{k}$-points 
for various values of $z$, along with values for 2H-TaSe$_2$ as given in
Ref. \onlinecite{Wei-arxiv}.  Remarkably, the {\em second} neighbor hopping integral $t_2$ 
is the dominant contribution for $z \leq 0.1263$.  Also notable is the strong dependence 
of first and third neighbor hoppings, $t_1$ and $t_3$, on the Nb-O distance, which causes
$t_1$ to become largest for $z$ somewhat greater than 0.1263.  The other parameters 
in the table are all relatively independent of the Nb-O distance.  The third and beyond 
interplane hoppings may appear small, but they must be compared keeping in mind that there 
are twice as many third and fourth interplane neighbors as first and second.  Since 
$t^\perp_1 \sim 2t^\perp_3 \sim 2t^\perp_4$, even these long distance hoppings are relevant.
Indeed, it is not possible to get good agreement in the band splittings at the points $\Gamma$,
M, K, and the minimum between K and $\Gamma$ without five interplane hopping parameters
(four for $z = 0.1167$).

The TB model for $z = 0.1167$ is strongly dominated by the second neighbor interactions 
between the Nb atoms, with other hoppings roughly an order of magnitude smaller.  
If $t_2$ were the only hopping parameter present, the system
would consist of three disjoint sub-lattices and thus states would be threefold degenerate
(although it would not be obvious on the standard band plot).  Including $t_1$ as a perturbation 
would couple them weakly and remove degeneracies.
In this case, the unit cell of the system could be taken as the second-neighbor lattice and the
Brillouin zone would be folded back so that the K(H) point in the band plots shown is mapped onto
the $\Gamma$(A) point.  The result of this folding back (and the splitting
that occurs because of non-zero $t_1$),  is evident in Fig. \ref{fig:Band1167}, 
in that the eigenenergy of the H point is 
nearly the same as for the A point.  The M(L) point is mapped onto itself, and the minimum between
$\Gamma$(A) and M(L) is mapped onto the K(H) point, which results in a band structure that looks
very much like a $\sqrt{3}$ larger triangular lattice with only 
nearest neighbor hopping. Similar 
dominance of 2nd neighbor hopping has been observed for 2H-TaSe$_2$.
\cite{Wei-arxiv}  Those authors relate their small value for
$t_1$ to phase cancellation of Wannier functions 
centered at first nearest neighbors, whereas in-phase contributions at second-neighbors
give a large $t_2$ matrix element.  

For $z = 0.1359$, we find that the nearest neighbor interaction becomes dominant.
From Table \ref{tbl:TightBinding} and Fig. \ref{fig:WF1359diff} we see that the 
increase in hybridization of the O anion $p$-state in LiNbO$_2$ as the Nb-O 
distance is decreased plays a significant role in first neighbor hopping, while at the
same time this hybridization is rather insignificant in second neighbor hopping. 
This behavior suggests that
a significant contribution to the hopping integrals comes from within the Nb planes, perhaps
through the hotdog structure noted in the previous section.
The band structure in this case looks  
similar to that of Li$_{0.5}$NbO$_2$, calculated by Novikov {\em et al.},\cite{Novikov94-2}
especially in the magnitude of the variations around the M and K points, relative to the
bandwidth.  Their result shows some splitting
of the bands occurring at the zone surface (along the lines A-L-H-A), which may 
arise from the two Nb sites becoming inequivalent due to the removal of one Li atom.
If the Nb sites were kept equivalent as the Li is removed, such as in the virtual crystal
approximation, then the similarities between the band structures suggest that 
the main impact on the electronic structure is captured by changing the Fermi energy
and allowing the O layers to move closer to the Nb layers, 
decreasing the Nb-O distance. 

Another point of interest we see in Figs. \ref{fig:Band1167} and \ref{fig:Band1359}
is a significant change in how the valence band is situated in relation to the
O $2p$ bands.  Nb $4d$ states in the conduction
band around the K point are significantly lowered in energy as the Nb-O distance is
decreased so that the bandgap becomes indirect and much smaller.  This change will
have contributions from both a Madelung shift and from changes arising from the
altered Nb-O interaction.

\begin{table*}[ht]
\begin{centering}
\begin{tabular}{|rc|ccc|ccc|ccc|}

\hline & Symmetry &\multicolumn{3}{c|}{Phonon frequency (meV)} &\multicolumn{3}{c|}{$\alpha$ for TO} &\multicolumn{3}{c|}{$\alpha$ for LO}	\\
\footnotetext[1]{IR active mode}
\footnotetext[2]{Raman active mode}
& Designation & Polarization & TO & LO 
& $\alpha_{\text{Li}}$ & $\alpha_{\text{Nb}}$ & $\alpha_{\text{O}}$& $\alpha_{\text{Li}}$ & $\alpha_{\text{Nb}}$ & $\alpha_{\text{O}}$ \\	
\hline\hline
1 					& E$_{2g}$ 	&	x-y	&	9.4		&		 	& 0    & 0.38 & 0.12 &&&\\
3 					& B$_{2g}$ 	&	z	&	18.5	&			& 0    & 0.39 & 0.11 &&&\\
\hline
4 					& E$_{2u}$ 	&	x-y	&	28.8	&			& 0.50 & 0    & 0.01 &&&\\
\footnotemark[1] 6  & E$_{1u}$ 	&	x-y	&	29.9	&	33.5	& 0.48 & 0.03 & 0    & 0.46 & 0.04 & 0 	  \\
\hline
\footnotemark[1] 8  & A$_{2u}$ 	&	z	&	53.3	&	70.0	& 0.47 & 0.01 & 0.03 & 0.43 & 0.06 & 0.01 \\
\footnotemark[2] 9  & E$_{2u}$ 	&	x-y	&	58.8	&			& 0.01 & 0    & 0.49 &&&\\
11					& E$_{1g}$ 	&	x-y	&	59.4	&			& 0    & 0    & 0.50 &&&\\
\footnotemark[1] 13 & E$_{1u}$	&	x-y	&	64.4	&	71.6	& 0    & 0.12 & 0.38 & 0.02 & 0.10 & 0.38 \\ 
15                  & B$_{1u}$ 	&	z	&	64.5	&			& 0.38 & 0    & 0.12 &&&\\
\footnotemark[2] 16 & E$_{2g}$	&	x-y	&	64.8	&			& 0    & 0.12 & 0.39 &&&\\
\hline
\footnotemark[2] 18 & A$_{1g}$  &	z	&	85.4	&			& 0    & 0    & 0.50 &&&\\
\footnotemark[1] 19 & A$_{2u}$  &	z	&	85.5	&	96.3	& 0.01 & 0.14 & 0.35 & 0.05 & 0.09 & 0.37 \\ 
20					& B$_{1u}$ 	&	z	&	89.5	&			& 0.13 & 0    & 0.38 &&&\\
21					& B$_{2g}$ 	&	z	&	94.8	&			& 0    & 0.11 & 0.39 &&&\\
\hline
\end{tabular}
\caption[Zone-center phonon frequencies]
{Calculated zone-center phonon frequencies, with LO modes listed where they occur.  See Appendix \ref{Apx:Phonon} for detailed descriptions of the atomic motions.  All phonons with x-y polarization are doubly-degenerate, as required by the hexagonal symmetry.  The last six columns show the calculated value
of $\alpha_i$, defined by equation \ref{eqn:PhononAlpha} for both TO and LO modes.}
\label{tbl:PhononFrequencies}
\end{centering}
\end{table*}

%
\section{Zone Center Vibrational Modes}

We have calculated the $3N-3 = 21$ optical phonon eigenmodes with $\mathbf{q} = 0$. To 
our knowledge, experimental IR or Raman measurements have not been reported on LiNbO$_2$.
However, we believe these results will be useful for comparison with electron-phonon
calculations done on the de-lithiated Li$_x$NbO$_2$, which will be explored in future work.
The frequencies are given in Table \ref{tbl:PhononFrequencies}.  Mode 
symmetries have been obtained with the use of SMODES, part of the {\em Isotropy} package 
of Stokes and Boyer.\cite{Stokes}  Masses of the ions used were 6.94 (Li), 92.91 
(Nb), and 16.00 (O), all in amu.  Also in Table \ref{tbl:PhononFrequencies}
are given values of the isotope shift $\alpha$, defined by
\begin{equation}
	\alpha_i = -\frac{\partial \ln \omega}{\partial \ln M_i},
	\label{eqn:PhononAlpha}
\end{equation}
where $i$ runs over the different (types of) atoms in the crystal.  The derivative
was approximated with a centered difference, which, for a harmonic crystal with $\omega \propto M^{-1/2}$, introduces errors of order $(\frac{\delta M}{M})^2$. Using $\delta M = 1$ amu, this amounts to an error of about 2\% for Li, and less
for heavier elements.  One would expect $\sum_i \alpha_i = \frac12$ for a harmonic mode.


The dielectric tensor for a hexagonal system such as this will be diagonal with two distinct elements.  For  frequencies much higher than phonon frequencies but well below the gap, the static dielectric constants, calculated without ionic contributions, are found to be $\epsilon_\infty^x = 10.3$ and $\epsilon_\infty^z = 3.78$.  One can then use the generalized Lyddane-Sachs-Teller relation, $\epsilon_0^\alpha = \epsilon_\infty^\alpha \prod_m (\omega^\alpha_{\text{LO},m} / \omega^\alpha_{\text{TO},m})^2$ (where $\alpha$ runs over the cartesian directions, and the product is taken over all IR active modes with polarization in that direction) to include the ionic contribution for the low frequency dielectric constant.  Doing so, we find for LiNbO$_2$ that $\epsilon_0^x = 15.9$ and $\epsilon_0^z = 8.27$.  It is typical for LDA calculations to overestimate somewhat the dielectric constants, presumably due to the band gap being underestimated.

The eigenmodes fall into groups which are easily distinguishable by frequency.  In the
following discussion we include degeneracy in the numbering of modes.  The first
group, modes 1-3, are low frequency modes involving displacements of massive Nb-O layers
against each other, with Li layers remaining fixed.
The second group, modes 4-7, involve the lighter Li layers, and lie around 29 meV.
The third group, modes 8-17 in the 53-65 meV range, involve motions 
where layers of atoms slide in the
$x-y$ plane against adjacent layers of atoms, or groups of layers bounce against each other
in the $z$ direction.  The last group of modes at 85-95 meV
involves layers nearby each other bouncing against
each other in the z direction.  
A more detailed discussion of the individual modes is given in Appendix \ref{Apx:Phonon}.





%
\section{Effective charges}

\begin{table}[ht]
\begin{centering}
\begin{tabular}{|r|ccc|ccc|}

\hline 						& \multicolumn{3}{c|}{LiNbO$_2$} & \multicolumn{3}{c|}{NaCoO$_2$} \\
\hline 						&	Li		& Nb		& O		 &  Na 		& Co 		& O		 \\
\hline $Z^*_{xx}$			&	1.10	& 2.26		& -1.68  &	0.87 	& 2.49		& -1.68	 \\
	   $Z^*_{zz}$			&   1.69    & 1.31      & -1.50  &  1.37	& 0.87		& -1.12  \\
	   $Z^{\text{Formal}}$  &   +1      & +3        & -2     &  +1 		& +3		& -2	 \\
\hline

\end{tabular}
\end{centering}
\caption[Effective Charges]{Calculated Born effective charges for LiNbO$_2$, compared with those
calculated for NaCoO$_2$ by Li, Yang, Hou and Zhu. \cite{Li-Yang}  Note the disparity in the effective charges for the alkali metals and the transition metal in the $z$-direction.  For O however, $Z^*$ is not far from isotropic.}
\label{tbl:Born}
\end{table}

Born effective charge calculations have been carried out as described by 
Gonze and Lee.\cite{Gonze-response-2}
The calculation was done using a planewave cutoff of 60 Ha and 
200 $\mathbf{k}$-points in the Brillouin zone to give 
effective charge tensors $Z^*$ for each atom which are diagonal, and 
obey $Z^*_{xx} = Z^*_{yy}$, as required by symmetry for a hexagonal crystal.
Results are given in Table \ref{tbl:Born} for the relaxed atomic structure. 
The effective charges of NaCoO$_2$ have been included in this table to illustrate the strong
similarity between these two systems, which are similarly layered transition metal oxides
but otherwise their electronic structures are quite different.

The effective charges of LiNbO$_2$ are rather different than those reported by Veiten and Ghosez\cite{linbo3} for LiNbO$_3$.  There are structural and chemical differences responsible for this; the structure of LiNbO$_3$ has Nb coordinated in a distorted octahedral environment and thus has low symmetry.  Perhaps more importantly, the formal charges of Nb are very different: Nb$^{3+}$ in LiNbO$_2$, but Nb$^{5+}$ in LiNbO$_3$, leaving the formal electronic configuration of Nb in LiNbO$_3$ as d$^0$.  The effective charges of such ions are known to sometimes be rather large, as is the case in LiNbO$_3$.

$Z^*_{xx}$(Li) is close to the formal charge of Li indicating ionic type response for in-plane motion.  In fact, all ions have effective charge closer to the formal value for in-plane displacement than for $\hat z$ displacement.  The charge tensor for Li shows similar anisotropy to that of Li in LiBC,\cite{Kwan-Woo} where it was inferred that Li is involved chemically in coupling between B-C layers.  Similarly here, it can be expected that Li has chemical involvement in interlayer coupling between the electron rich Nb-O layers more so than Na does in Na$_x$CoO$_2$.  This involvement is not strong enough to inhibit de-intercalation, however.

The smaller than formal charges for Nb 
indicates substantial covalent character to the bonding, which is especially strong 
in the $\hat z$ direction.  This charge `renormalization' may be related to the strong change
in band structure due to change of the distance between the Nb and O layers, discussed in 
earlier sections.

\section{Summary}

In this work we have examined the electronic structure, the Wannier functions of the 
valence bands and their tight-binding parametrization, 
and several vibrational properties of the quasi-two-dimensional material LiNbO$_2$.  
These basic properties are important in providing the basis for understanding the 
behavior of the system when it is hole-doped by de-intercalation of Li.

Li$_x$NbO$_2$ seems to be a candidate for an insulator-to-superconductor
transition, as it undergoes an insulator-to-metal transition
and becomes superconducting, but doping studies are not systematic enough yet to shed light on this possibility.
Phase space considerations in two-dimensional materials lead to superconductivity 
that is independent of doping level, at least when the Fermi level is not far from a
band edge.\cite{role2D}  This property supports the possibility of an insulator-superconductor transition
with doping, i.e. as soon as it becomes metallic it is superconducting.

The Wannier function, centered on and based on the occupied Nb $d_{z^2}$ orbital, illustrates
graphically not only how Nb bonds with neighboring O ions, but also that it couples with
oxygens in adjoining layers.  Inspection of the $\mathbf{k}$-space wavefunction suggests there are
direct Nb-Nb interactions within the d$_{z^2}$ band as well as indirect coupling through O ions.

A tight-binding representation of the band, obtained by direct calculation (rather than fitting) using the Wannier function, reveals that several neighbors both in-plane and inter-plane are necessary to reproduce the dispersion.  To model the interlayer interaction precisely requires four to five interlayer hopping parameters.  For the experimental structure intralayer hopping is dominated by 2nd neighbor hopping, but the strength of both 1st and 3rd neighbor hopping is strongly modulated by varying the Nb-O distance.  This electron-lattice coupling provides one possibility for the superconducting pairing mechanism in Li$_x$NbO$_2$.

Comparison with previous calculations of the band structure of
Li$_{0.5}$NbO$_2$ suggests that the electronic structure of Li$_x$NbO$_2$ can be modeled in
virtual crystal fashion, or perhaps even in rigid band fashion if that is convenient.
The Nb $d_{z^2}$ band is
remarkably well isolated from neighboring bands by Madelung shifts and crystal field splitting.
We suspect that the overriding importance of this system will result from its unique
standing as a single band, triangular lattice system for which sophisticated studies on
simple models may be directly comparable to experimental data.  

\section{Acknowledgments}
We have benefited from communication with K. E. Andersen and A. Simon.
Some of the calculations of LiNbO$_2$ wavefunctions used to compute tight-binding parameters
were done on machines at Lawrence Livermore National Laboratory.
This work was supported by National Science Foundation Grant
DMR-0421810.

\appendix

\section{Phonon mode descriptions}
\label{Apx:Phonon}

Following is a descriptive list of the atomic motions for each mode given in table \ref{tbl:PhononFrequencies}, grouped by their spectroscopic activity.  Degeneracy is included in the numbering of modes. 

\subsection{Silent Modes}
Modes 1-2:  NbO$_2$ layers slide against each other in the x-y plane.

Mode 3:  NbO$_2$ layers bounce against each other in the z direction.

Modes 4-5:  LiO$_2$ layers slide against each other in the x-y direction.  Li displacements are roughly an order of magnitude larger than O displacements, which results in the mass of the O
atoms having very little effect on the frequency of this mode.

Modes 11-12:  Adjacent O layers slide against each other, with Li and Nb layers remaining 
stationary.

Mode 15:  LiO$_2$ layers bounce against each other in the z-direction.  This is the one high frequency mode whose isotope shift $\alpha$ gets its major contribution from the Li atoms.

Mode 20:  Out-of-phase Li layers beat against adjacent O layers in the z direction.  

Mode 21:  Out-of-phase Nb layers beat against adjacent O layers in the z direction.  

\subsection{IR Active Modes}

Mode 6-7:  In-phase Li layers sliding against NbO$_2$ layers.  The main contribution to $\alpha$ comes from the light Li atoms, with the heavy NbO$_2$ layers making very little contribution.  This mode is also IR active, with a small LO-TO splitting $\Delta \omega = \omega_{\text{LO}} - \omega_{\text{TO}} = 3.6$ meV.

Mode 8:  The TO mode consists of Li layers bouncing against NbO$_2$ layers in the z direction, and can be thought of as the ``z-polarized version'' of modes 6 and 7.  When the macroscopic electric field is included, the O displacements change sign and the mode becomes Li-O layers bouncing against Nb layers.  $\Delta \omega = 16.7$ meV.

Modes 13-14:  Li and Nb are in synchronized sliding against O layers.  The metal layers are all in phase with each other, and out of phase with the O layers.  This mode is IR active, with $\Delta \omega = 7.2$ meV.

Mode 19:  The transverse mode consists of LiO$_2$ layers beating against Nb layers in the z-direction.  The Li-O layers are all in phase.  This mode is IR active, and once the macroscopic electric field is included, this mode can be characterized as having Nb layers bounce against adjacent O layers, where the Nb layers are out of phase.  $\Delta \omega = 10.8$ meV.

\subsection{Raman Active Modes}

Modes 9-10:  Out-of-phase Li layers slide in the x-y direction against adjacent O layers.  The rest of the phonon modes from here on have the contribution to their isotope shift $\alpha$ dominated by O.

Modes 16-17:  Out-of-phase Nb layers slide against adjacent O layers in the x-y plane.

Mode 18:  The O layers beat against each other in the z-direction.  This fully symmetricRaman active mode is of particular interest, as it corresponds to the variation of the internal structural parameter $z$.  This mode has a significant impact on the electronic structure as discussed in previous sections, so we suspect that it may have particularly strong electron-phonon coupling in hole-doped Li$_x$NbO$_2$.

\subsection{Clarifying Comments}

The $z$-polarized A$_{2u}$ IR active modes 8 and 19 have their
eigenvectors affected by the inclusion of the macroscopic electric field.
Without the electric field, the eigenmodes can be described as one where Nb layers bounce
against Li-O units  and 
one where  Li layers bounce against Nb-O units.  The application of the electric field causes
these vectors to mix, and the descriptions change so that the latter one is
replaced by a mode where Li and Nb bounce against O 
layers.  The distinction between these two modes is the relative sign for Li 
and Nb displacements (negative for the former, positive for the latter.)  It is not
unusual to have modes mix to give different LO modes, as the 
non-analytic part of the dynamical matrix which incorporates the 
electric field can have eigenvectors differing from the analytic part of the 
dynamical matrix, resulting in a mixing of modes of the same symmetry.  
This is discussed in some detail by Gonze and Lee.\cite{Gonze-response-2}
This complicates
the association of LO modes with their corresponding TO mode, so we have chosen to
pair the LO and TO modes such that $\omega_{\text{LO}} > \omega_{\text{TO}}$.

\bibliographystyle{apsrev}
\bibliography{LiNbO2}

\end{document}